\documentclass[aip,jcp,reprint,noshowkeys,superscriptaddress]{revtex4-1}
\usepackage{subcaption}
\usepackage{bm,graphicx,tabularx,array,booktabs,dcolumn,xcolor,microtype,multirow,amscd,amsmath,amssymb,amsfonts,physics,siunitx,enumitem}
\usepackage[version=4]{mhchem}
\usepackage[utf8]{inputenc}
\usepackage[T1]{fontenc}
\usepackage{txfonts}
\usepackage[normalem]{ulem}
\usepackage{mleftright}
\usepackage{wrapfig}

\AtBeginDocument{\nocite{achemso-control}}

\usepackage{tikz}
\usetikzlibrary{arrows.meta,positioning,shapes.misc,arrows}

\definecolor{hughgreen}{RGB}{100, 0, 140}

\usepackage{titlesec}
\titlespacing\section{10pt plus 2pt minus 2pt}{10pt plus 2pt minus 2pt}{10pt plus 2pt minus 2pt}
\titlespacing\subsection{2pt plus 2pt minus 2pt}{10pt plus 2pt minus 2pt}{5pt plus 2pt minus 2pt}
\titlespacing\subsubsection{2pt plus 2pt minus 2pt}{10pt plus 2pt minus 2pt}{5pt plus 2pt minus 2pt}

\allowdisplaybreaks

\usepackage[
	colorlinks=true,
    citecolor=blue,
    linkcolor=blue,
    filecolor=blue,      
    urlcolor=blue,	
    breaklinks=true
	]{hyperref}
\newcommand{\etal}{\textit{et al.}~}

\newcommand{\ceri}[2]{\mleft(#1|#2\mright)} 

\newcommand{\hH}{\Hat{H}} 
\newcommand{\hS}{\Hat{S}} 
\newcommand{\hR}{\Hat{R}}

\newcommand{\hE}{\Hat{E}} 
\newcommand{\cre}[1]{\hat{a}_{#1}^\dagger} 
\newcommand{\ani}[1]{\hat{a}_{#1}^{\vphantom{\dagger}}} 

\newcommand{\e}{\mathrm{e}}
\newcommand{\Eh}{\mathrm{E_h}}
\newcommand{\s}{\mathrm{s}}
\newcommand{\tS}{\mathrm{S}}
\newcommand{\tT}{\mathrm{T}}

\usepackage{upgreek}
\newcommand{\sigg}{{\upsigma_\textrm{g}}}
\newcommand{\sigu}{{\upsigma_\textrm{u}}}

\newcommand{\Nbas}{n}
\newcommand{\Ndet}{M}

\newcommand{\br}{\bm{r}}
\newcommand{\bx}{\bm{x}}
\newcommand{\bR}{\bm{R}}
\newcommand{\bS}{\bm{S}}

\newcommand{\hd}{\hphantom{-}}

\newcommand{\UOX}{Physical and Theoretical Chemical Laboratory, Department of Chemistry, University of Oxford, Oxford, OX1 3QZ, U.K.}
\newcommand{\LPCQ}{Current address: Laboratoire de Chimie et Physique Quantiques (UMR 5626), Universit\'{e} de Toulouse, CNRS, UPS, Toulouse, France}

\begin{document}	

\title{Excited states, symmetry breaking, and unphysical solutions on the CASSCF energy landscape}
\title{Excited states, symmetry breaking, and unphysical solutions in state-specific CASSCF theory}

\date{\today}
\author{Antoine \surname{Marie}}
\affiliation{\UOX}
\affiliation{\LPCQ}
\author{Hugh G.~A.~\surname{Burton}}
\email{hgaburton@gmail.com}
\affiliation{\UOX}

\begin{abstract}
\begin{wrapfigure}[10]{r}{0.38\textwidth}
    \flushleft
    \vspace{-0.4cm}
    \hspace{-1.7cm}
    \fbox{\includegraphics[width=0.38\textwidth]{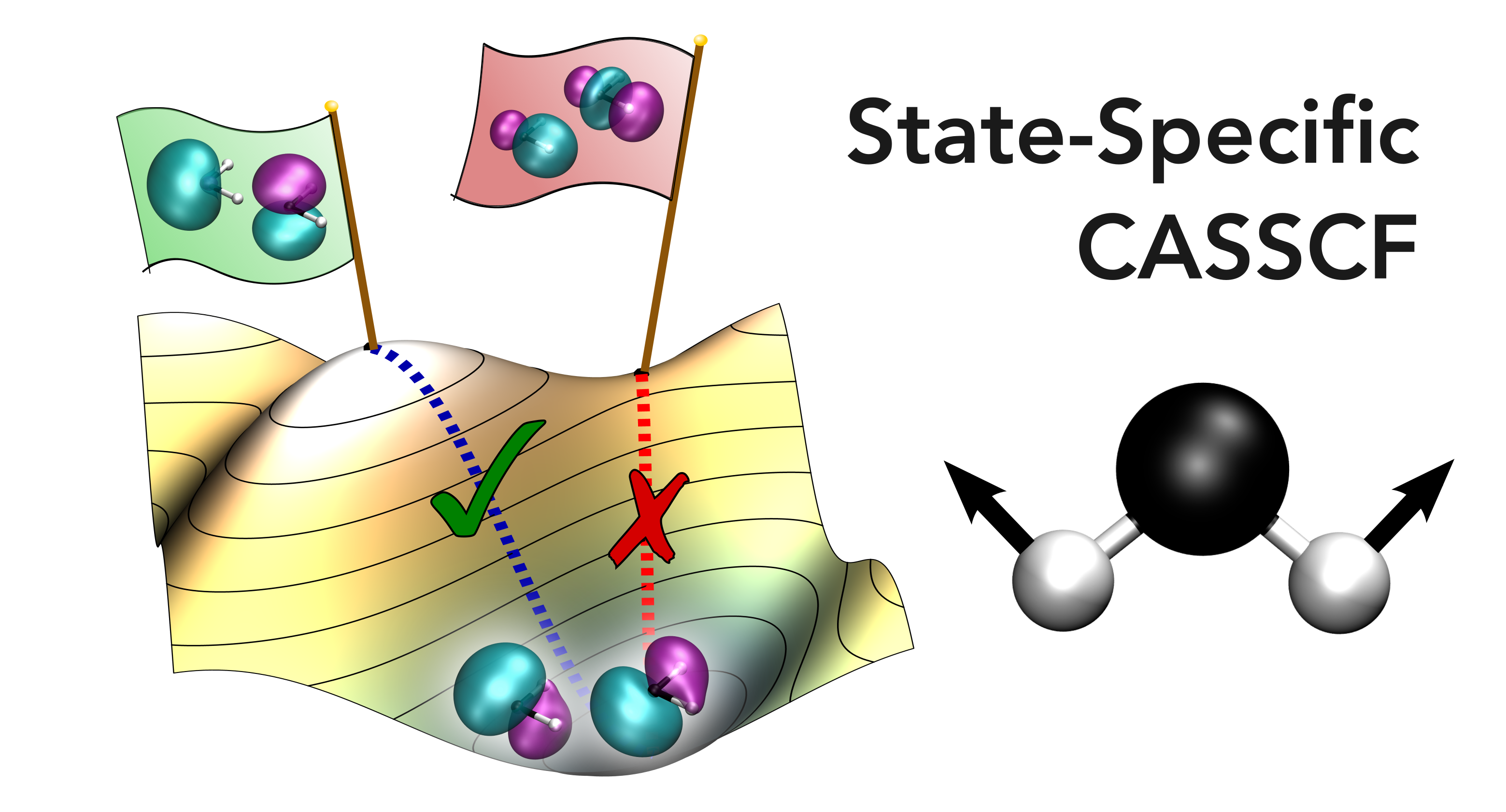}}
\end{wrapfigure}
State-specific electronic structure theory provides a route towards balanced excited-state 
wave functions by exploiting higher-energy stationary points of the electronic energy. 
Multiconfigurational wave function approximations can describe both
closed- and open-shell excited states and avoid the issues associated with state-averaged approaches.
We investigate the existence of higher-energy solutions in complete active space 
self-consistent field (CASSCF) theory and characterise their topological properties.
We demonstrate that state-specific approximations can provide accurate higher-energy excited states 
in \ce{H2} (6-31G) with more compact active spaces than would be required in a state-averaged formalism.
We then elucidate the unphysical stationary points, demonstrating that they arise from 
redundant orbitals when the active space is too large, or symmetry breaking when the active space is 
too small.
Furthermore, we investigate the conical intersection in \ce{CH2} (6-31G) and the avoided crossing
in \ce{LiF} (6-31G), revealing the severity of root flipping and demonstrating that state-specific
solutions can behave quasi-diabatically or adiabatically.
These results elucidate the complexity of the CASSCF energy landscape, highlighting the advantages 
and challenges of practical state-specific calculations.
\end{abstract}

\maketitle
\linepenalty1000
\raggedbottom

\section{Introduction}
\label{sec:intro}


Electronic excited states are fundamentally higher-energy solutions to the time-independent 
Schr\"{o}dinger equation. 
``State-specific'' representations can be identified using 
higher-energy stationary points of the electronic energy landscape.\cite{Burton2022a}
The exact excited states in full configuration interaction (FCI) correspond to energy 
saddle points and the number of downhill Hessian eigenvalues increases with each energy 
level.\cite{Olsen1982,Golab1983,Olsen1983,Golab1985,Burton2022a,Bacalis2016,Bacalis2020}
Higher-energy stationary points also exist in non-linear wave function approximations,
but the development of practical state-specific methods has been hindered by the 
challenges of non-ground-state optimisation, the non-linearity of the electronic energy
landscape, and the presence of unphysical solutions.

Instead, the workhorse of modern excited-state electronic struture theory is linear-response time-dependent
density functional theory (LR-TDDFT), which predicts excitation energies from the response
of the ground-state electron density to a weak external perturbation.\cite{Runge1984,Dreuw2005,Burke2005}
Despite its computational efficiency, LR-TDDFT inherits the failures of approximate Kohn-Sham (KS) 
exchange-correlation functionals, creating large errors for bond dissociation or open-shell electronic states.\cite{Hait2019}
Furthermore, the ubiquitous adiabatic approximation excludes double excitations and their associated 
avoided crossings.\cite{Burke2005,Maitra2004}
Alternative single-reference methods, such as algebraic diagrammatic construction\cite{Schirmer1982,Dreuw2015} 
(ADC) and equation-of-motion coupled cluster\cite{Stanton1993,Krylov2008} (EOM-CC) 
can provide more accurate excitation energies at a greater
computational cost, but depend strongly on the quality of the reference determinant.
The strong influence of the ground-state orbitals can also create an unbalanced description of charge transfer 
and Rydberg excitations,\cite{Tozer2003,Dreuw2004}
 where significant electronic relaxation can occur.\cite{McLachlan1964,Runge1984,Dreuw2005,Burke2005,Bartlett2012,HelmichParis2019}

These challenges have encouraged 
researchers to revisit excited state-specific approximations.
For higher-energy SCF calculations ($\Delta$SCF), 
this progress has been catalysed by the development of new optimisation algorithms 
that avoid variational collapse to the ground state, including the maximum overlap method,\cite{Gilbert2008,Barca2014,Barca2018} 
square-gradient optimisation,\cite{Hait2021,Hait2020} state-targeted energy projection,\cite{CarterFenk2020} 
quasi-Newton direct orbital optimisation,\cite{Levi2020,Levi2020a,Ivanov2021}
and generalised variational principles.\cite{Shea2017}
Recent calculations have shown that higher-energy Hartree--Fock (HF) and KS-DFT solutions 
can accurately describe charge transfer and double excitations at a low computational cost.\cite{Gilbert2008,Hait2020}
Beyond SCF approximations, higher-energy variational or projective coupled-cluster ($\Delta$CC) solutions
can provide more accurate double  and double-core excitations by incorporating dynamic electron 
correlation.\cite{Jankowski1994,Jankowski1994a,PiecuchBook,Mayhall2010,Lee2019,Kossoski2021,Marie2021a} 
%
While $\Delta$SCF and $\Delta$CC are successful for double and charge transfer excitations, 
these single-reference methods cannot describe open-shell excited states and statically correlated ground states.
The onset of this failure usually becomes apparent through spin contamination,\cite{Burton2021a,Burton2018}
spontaneous symmetry breaking,\cite{Coulson1949,Fukutome1973,Fukutome1974,Fukutome1974a,Burton2021a,Barca2014,Hait2019,Ye2017}
and additional unphysical 
solutions.\cite{Burton2021a,Jankowski1994,Jankowski1994a,PiecuchBook,Mayhall2010,Kossoski2021,Marie2021a}
Furthermore, the solutions of interest can disappear as the molecular structure changes, 
creating discontinuous excited-state energy surfaces or 
gradients.\cite{Thom2008,Marie2021a,Burton2021a,Burton2020,Jensen2018,Vaucher2017,Dong2020}

Multiconfigurational SCF (MCSCF) methods,\cite{SzaboBook} particularly the 
complete-active space self-consistent field (CASSCF) formulation,\cite{Das1966,Roos1980a,Roos1980b}
are the state-of-the-art for describing statically correlated electronic systems.\cite{RoosBook}
The CASSCF wave function is a linear expansion of all the configurations
that can be constructed from a set of partially occupied ``active orbitals'',
and the energy is optimised with respect to the configuration interaction (CI) and orbital coefficients simultaneously.\cite{Roos1980a} 
It has long been known that higher-energy MCSCF solutions can represent electronic 
excited states,\cite{Das1973,Krauss1976,Bauschlicher1978,Bauschlicher1980a,Bauschlicher1980b,Bauschlicher1980c}
and that multiple symmetry-broken CASSCF solutions can occur for an inadequate active space.\cite{Guihery1997,Angeli2003}
More recently, MCSCF expansions truncated to single excitations have shown 
promise for singly excited charge transfer states,\cite{Shea2018,Shea2020,Zhao2020a,Zhao2020b,Hardikar2020}
while state-specific configuration interaction with higher degrees of truncation can 
handle challenging multireference problems, singly-, and doubly-excited states.\cite{Kossoski2022a}
However, the strong coupling between the orbital and CI degrees of freedom makes the optimisation
challenging, and second-order optimisation algorithms are generally required to reach 
convergence in practice.%
\cite{Dalgaard1978,Dalgaard1979,Yeager1979,Lengsfield1980,Seigbahn1981,Werner1981,Werner1985,Yeager1980a,Yeager1980b,Jorgensen1981,Yeager1982,Sun2017,Kreplin2019,Kreplin2020}

Extensive research in the 1980s focused on characterising higher-energy MCSCF solutions.
It was originally suggested that an $n$th excited state approximation should be the 
$n$th state in the configuration expansion.\cite{Lengsfield1980}
However, this requirement is often not achieved, resulting in ``root flipping''.\cite{Das1973,Werner1981,Olsen1982}
Furthermore, several stationary points satisfying this condition can often be identified.\cite{Golab1983,Golab1985,Rizzo1990}
The enormous complexity of the multiconfigurational solution space led Golab \etal{}\ to conclude that 
\textit{``selecting an MCSCF stationary point is a very severe problem.''}\cite{Golab1983}
Instead, the state-averaged (SA) approach is generally used, where a weighted average energy of the 
$n$ lowest CI states constructed from one set of orbitals is optimised.\cite{Werner1981}
While this approach has become the method of choice for excited-state CASSCF, it has several disadvantages:
discontinuities can occur on the SA-CASSCF potential energy surface if two states
require orbitals with significantly different character;\cite{Zaitsevskii1994}
the number of states is limited by the size of the active space; large active spaces
are required to target high-lying states; and the Helmann-Feynamn theorem 
cannot be applied to compute nuclear gradients because individual
SA-CASSCF solutions are not stationary points of the energy.

Recently, the limitations of SA-CASSCF and the development of non-ground-state SCF optimisation
algorithms has inspired several new investigations into state-specific CASSCF excited states. 
In particular, Neuscamman and co-workers have developed generalized 
variational principles\cite{Tran2020,Hanscam2022} and the $W\Gamma$ approach inspired by MOM-SCF,\cite{Tran2019}
demonstrating that the issues of root flipping and variational collapse to the ground state 
can be successfully avoided. 
Despite these advances, we still do not have a complete understanding of the multiple stationary points on the 
SS-CASSCF energy landscape and several practical questions remain.
For example, how many stationary points are there and how does this change with the 
active space or basis set size?
Where do unphysical solutions arise, what are their characteristics, and
when does symmetry breaking occur?
And finally, do state-specific excitations behave diabatically or adiabatically as the molecular structure 
evolves?

Our aim in this work is to answer these questions and establish
a theoretical foundation for practical excited state-specific calculations.
Using second-order optimisation techniques, we investigate the existence and properties 
of multiple CASSCF solutions in typical molecular systems.
Our numerical optimisation exploits analytic gradients and second derivatives of the CASSCF energy,
and the relevant differential geometry is summarised below.
Using these techniques, we comprehensively enumerate the multiple CASSCF solutions in 
\ce{H2} (6-31G) and characterise the resulting unphysical solutions.
We find that state-specific calculations can accurately describe high-lying excitations
with fewer active orbitals than state-averaged formalisms, 
and reveal that multiple solutions can arise from active spaces that are too large or too small.
We then investigate the conical intersection in \ce{CH2} (6-31G) and the 
avoided crossing of \ce{LiF} (6-31G), demonstrating the importance and difficulty
of selecting the correct physical solution.

\section{Exploring the multiconfigurational energy landscape}
\label{sec:theoretical}

\subsection{Defining the CASSCF wave function}
\label{subsec:CASSCF}
A multiconfigurational wave function is defined as the linear combination of $\Ndet$ Slater determinants
\begin{equation}
\ket{\Psi_k} = \sum_{I=1}^{\Ndet} C_{I k} \ket{\Phi_I},
\label{eq:CIwfn}
\end{equation}
where $\ket{\Phi_I}$ represents different configurations built from a common set of 
molecular orbitals (MO) $\phi_p(\bx)$ and the $C_{Ik}$ are the variable 
CI coefficients for state $k$.\cite{HelgakerBook}
Here, $\bx = (\br,\sigma)$ is the combined spatial and spin electronic coordinate.
The MOs are constructed as linear combinations of $\Nbas$ (nonorthogonal) atomic orbitals (AO) $\chi_{\mu} (\bx)$ as
\begin{equation}
  \label{eq:mo}
  \phi_p(\bx) = \sum^{\Nbas}_\mu \chi_{\mu}(\bx)\, c_{\cdot p}^{\mu \cdot},
\end{equation}
where we use the nonorthogonal tensor notation of Ref.~\onlinecite{HeadGordon1998} and the 
$c_{\cdot p}^{\mu \cdot}$ denote the variable MO coefficients.
Normalisation of the wave function, and orthogonalisation of the MOs, is guaranteed by the constraints
\begin{equation}
\sum_{I=1}^{\Ndet} \abs{C_I}^2 = 1
\quad\text{and}\quad
\sum_{\mu=1}^{\Nbas}  (c^*)_{p \cdot}^{\cdot \mu}\, \braket*{\chi_\mu}{\chi_\nu}\, c_{\cdot q}^{\nu \cdot} = \delta_{pq},
\label{eq:constraint}
\end{equation}
where $\braket*{\chi_\mu}{\chi_\nu}$ denotes the AO overlap matrix elements.
We will only consider wave functions where $C_{Ik}$ and $c_{\cdot p}^{\mu \cdot}$ are real. 

When every electronic configuration for a finite basis set is included in an FCI expansion, 
the global minimum on the parametrised 
electronic energy landscape corresponds to the exact ground state.\cite{Burton2022a}
Excited states form saddle points of the energy and the number of downhill directions 
increases with each excitation.\cite{Burton2022a,Bacalis2016,Bacalis2020,Golab1983}
The FCI wave function is invariant to unitary transformations of the MOs, but 
the number of configurations scales exponentially with the system size.

The complete active space (CAS) framework builds a truncated expansion using every configuration
within a set of ``active orbitals'' that describe the dominant static electron correlation.\cite{Roos1980a}
The orbitals are partitioned into inactive and virtual orbitals that are doubly occupied or empty in every 
configuration, respectively, and active orbitals with varying occupations.
Simultaneously optimising the energy with respect to the orbital and CI coefficients leads to the state-specific CASSCF 
approach and gives true stationary points of the electronic energy.\cite{Roos1980a,Roos1980b,Seigbahn1981}
If the CASSCF wave function targeting the $k$th excited state is represented by the $k$th eigenstate of the 
corresponding CAS-CI expansion, then the Hylleraas-Undheim-MacDonald theorem\cite{Hylleraas1930b,MacDonald1933}
also provides a upper bound to the excited-state energy.\cite{Golab1983}

\subsection{Differential geometry of the CASSCF energy}
\label{subsec:differentialGeometry}
We exploit an exponential form of the CASSCF wave function that conserves the 
orthogonality constraints [Eq.~\eqref{eq:constraint}].\cite{Dalgaard1978,Yeager1979}
Starting from an initial CASSCF wave function $\ket{\Psi_0}$, an arbitrary step 
can be defined using unitary transformations as
\begin{equation}
  \label{eq:mcscf_wf}
  \ket{\Psi}=\e^{\hR} \e^{\hS}\ket{\Psi_0},
\end{equation}
where $\e^{\hR}$ and $\e^{\hS}$ account for orbital relaxation and transformations of
the CI component, respectively.
The $\hR$ operator is anti-Hermitian and is defined using the second-quantised 
creation and annihilation operators for the current MOs as\cite{Dalgaard1978,Douady1980}
\begin{equation}
\hR = \sum_{p>q} R_{pq}\, \hE^{-}_{pq},
\label{eq:Rop}
\end{equation}
where the spin-adapted one-body anti-Hermitian replacement operators are\cite{HelgakerBook}
\begin{equation}
\hE^{-}_{pq} = \sum_{\sigma \in \{\uparrow, \downarrow \} }  \cre{q\sigma}\ani{p\sigma} - \cre{p\sigma}\ani{q\sigma}.
\end{equation}
The invariance of the energy with respect to inactive-inactive, active-active, 
and virtual-virtual orbital transformations
means that $R_{pq}$ can be further restricted to only excitations between different
sub-blocks.
Similarly, $\e^{\hS}$ performs a unitary transformation between the CI 
component of $\ket{\Psi_0}$ and the remaining orthogonal states $\ket{\Psi_{K}}$ 
in the current CASCI space, with $\hS$ defined as\cite{Yeager1979}
\begin{equation}
\hS = \sum_{K \neq 0} S_{K} \Big( \ket{\Psi_K}\bra{\Psi_0}-\ket{\Psi_0}\bra{\Psi_K} \Big).
\label{eq:Sop}
\end{equation}

Using the exponential parametrisation, the CASSCF energy can be expressed as 
\begin{equation}
\label{eq:casscf_energy}
E(\bR,\bS) = \mel{\Psi_0}{\e^{-\hS} \e^{-\hR} \hH \e^{\hR} \e^{\hS}}{\Psi_0},
\end{equation}
where $\bR$ and $\bS$ are vectors that gather the $R_{pq}$ and $S_K$ coefficients
in the orbital and CI transformations, respectively, and $\hH$ is the electronic Hamiltonian.
Stationary points of $E$, corresponding to optimal CASSCF solutions, then occur when 
the gradients with respect to orbital and CI transformations are simultaneously zero.
Performing a Baker--Campbell--Hausdorff expansion of the energy to 
second order gives\cite{Yeager1979}
\begin{equation}
\begin{split}
E \approx &\mel*{\Psi_0}{\hH}{\Psi_0} 
+ \mel*{\Psi_0}{[\hH, (\hR + \hS)]}{\Psi_0}
\\
&+ \frac{1}{2} \mel*{\Psi}{[[\hH,(\hR + \hS)],(\hR + \hS)]}{\Psi_0} + \dots
\end{split}
\end{equation}
Expressions for the first- and second-derivates of the energy 
can then be identified as
\begin{subequations}
\begin{align}
\left.\frac{\partial E}{\partial R_{pq}}\right|_{\bR,\bS=\bm{0}} &= \mel*{\Psi_0}{[\hH, \hE^{-}_{pq}]}{\Psi_0},
\label{eq:gradOrb}
\\
\left.\frac{\partial E}{\partial S_{K}}\right|_{\bR,\bS=\bm{0}} &= 2 \mel*{\Psi_0}{\hH}{\Psi_K},
\label{eq:gradCI}
\end{align}
\end{subequations}
and
\begin{subequations}
\begin{align}
\left.\frac{\partial^2 E}{\partial R_{pq}\partial R_{rs}}\right|_{\bR,\bS=\bm{0}} 
&= \frac{1}{2}(1+P_{pq,rs}) \mel*{\Psi_0}{[[\hH, \hE^{-}_{pq}],\hE^{-}_{rs}]}{\Psi_0},
\\
\left.\frac{\partial^2 E}{\partial R_{pq}\partial S_{K}}\right|_{\bR,\bS=\bm{0}} 
&= 
\mel*{\Psi_0}{[\hH, \hE^{-}_{pq}]}{\Psi_K},
\\
\left.\frac{\partial E}{\partial S_{L} \partial S_{K}}\right|_{\bR,\bS=\bm{0}} 
&= 2 \mel*{\Psi_K}{\hH - E_0}{\Psi_L},
\end{align}
\end{subequations}
where $E_0$ is the energy at $\bR, \bS = \bm{0}$, $P_{pq,rs}$ permutes the $(pq)$ and $(rs)$ indices, 
and the Hermiticity of $\hH$ and $[\hH, \hE_{pq}^{-}]$ have been exploited.
Explicit formulae for these expressions have been summarised elsewhere [see Ref.~\onlinecite{Olsen1983}]
but are given in the Supporting Information (Section~S1) for completeness.

Note that 
the first  and second derivatives can only be computed 
when $\bR = \bm{0}$ and $\bS = \bm{0}$.\cite{Douady1980}
Therefore, after taking a step in the parameter space, the energy gradient and Hessian 
must be computed using the new MOs and CI vectors corresponding to the updated wave function.
A similar shift in the reference state after each step is also required for second-order 
HF optimisation algorithms.\cite{Voorhis2002,Burton2021a}

\subsection{Characterising distinct solutions}
\label{subsec:distinguishSolutions}
The invariance to unitary transformations within each orbital partition
means that the same CASSCF wave function can be identified with different CI or MO coefficients.
We use the overlap between two stationary solutions $\ket{^x\Psi}$ and $\ket{^w\Psi}$ 
to define a positive semidefinite distance metric
\begin{equation}
d(x,w) = 1 - \abs{\braket*{^x\Psi}{^w\Psi}\,}.
\end{equation}
The overlap for two arbitrary CI wave functions with $\Ndet_x$ and $\Ndet_w$
configurations, respectively, is given by
\begin{equation}
\braket*{^x\Psi}{^w\Psi} 
= 
\sum_{I=1}^{\Ndet_x} \sum_{J=1}^{\Ndet_w} \,^{x}C^{*}_{I} \braket*{^x\Phi_I}{^w\Phi_J} \,^{w}C_{J}.
\end{equation}
Since $\ket{^x\Psi}$ and $\ket{^w\Psi}$ have different sets of MOs, evaluating the
overlap matrix elements $\braket*{^x\Phi_I}{^w\Phi_J}$ requires a nonorthogonal framework.
We compute these matrix elements using the extended nonorthogonal Wick's theory,\cite{Burton2021c,Burton2022c} 
which avoids the computationally expensive generalized Slater--Condon rules.\cite{MayerBook}

To understand the MOs in a CASSCF solution, we canonicalise the inactive and virtual
orbitals and construct natural orbitals within the active space.
The canonical inactive and virtual orbitals, and their associated orbital energies,
are identified by diagonalising the relevant sub-blocks of the 
Fock matrix, defined as\cite{HelgakerBook}
\begin{equation}
F_{pq} = h_{pq} + \sum_{rs} \gamma_{rs}\qty(\ceri{pq}{sr} - \frac{1}{2}\ceri{pr}{sq}).
\label{eq:general_fock}
\end{equation}
Here, $\gamma_{pq}$ denotes the one-body reduced density matrix elements
in the MO basis, $h_{rq}$ are the one-electron Hamiltonian matrix elements, and $\ceri{pq}{rs}$ are
the two-electron repulsion integrals.
The natural orbitals within the active space are the eigenvectors of the one-body reduced 
density matrix and their eigenvalues are the occupation numbers $n_p$.\cite{Lowdin1955a}

\subsection{Optimization techniques}
\label{subsec:optimization}
Since we are concerned with understanding the CASSCF solution space, we require an algorithm 
capable of converging arbitrary stationary points on the energy landscape, including minima and 
higher-index saddle points.
Higher-energy CASSCF stationary points are notoriously difficult to converge due to the strong coupling 
between the orbital and CI degrees of freedom,\cite{Das1973,Yeager1979,Bauschlicher1980b,Bauschlicher1980c,Yeager1980a}
and the possibility of root flipping in the configuration space.\cite{Docken1972,Werner1981}
Therefore, we employ second-order techniques that introduce the orbital-CI coupling through 
the analytic Hessian matrix of second derivatives.
These algorithms are too computationally expensive to be practical for larger systems,
but they are sufficient for understanding the CASSCF solutions in small molecules.

We search for multiple solutions using several initial guesses generated using random
orbital and CI rotations from the ground state HF solution.
The eigenvector-following technique with analytic gradient and Hessian information was used to
target stationary points with a particular Hessian index.\cite{Cerjan1981,Wales1990} 
While this method has been described in detail elsewhere [see Ref.~\onlinecite{WalesBook}],
we include a summary in the Supporting Information (Section~S2).
Related mode-following methods have previously been applied to locate higher-energy 
electronic stationary points in multiconfigurational\cite{Olsen1982,Golab1983,Olsen1983,Hoffmann2002}
and single-determinant\cite{Burton2021a} SCF calculations.
The convergence behaviour was further improved with a modified trust region approach based
on the dogleg method.\cite{NocedalBook} 
Trust region methods are a well-established approach for controlling the convergence
of second-order methods in CASSCF calculations.\cite{Yeager1980a,Yeager1980b,Jorgensen1981,Yeager1982}
Once a set of stationary points have been identified, their evolution with changes 
in the molecular structure can be determined 
by using the optimised orbital and CI coefficients at one geometry 
to define an initial guess at the next geometry.
Since the Hessian index may not be conserved along a reaction coordinate,\cite{Olsen1982}
these subsequent calculations are performed using a trust region Newton--Raphson 
algorithm, as described in the Supporting Information (Section~S3).

We have implemented this numerical optimisation in an extension to the
PySCF software package.\cite{PySCFb}
The convergence threshold for the root-mean-squared value of the gradient amplitudes was 
universally set to $10^{-8}\,\Eh$.
The canonical and natural orbitals for stationary points were 
subsequently computed using PySCF and visualised using VMD.\cite{VMD} 
All other graphical figures were created using Mathematica~12.0.\cite{Mathematica}


\section{Results and Discussion}

\subsection{Molecular \ce{H2} dissocation}
\label{sec:results}

\begin{figure*}
\includegraphics[width=0.88\linewidth]{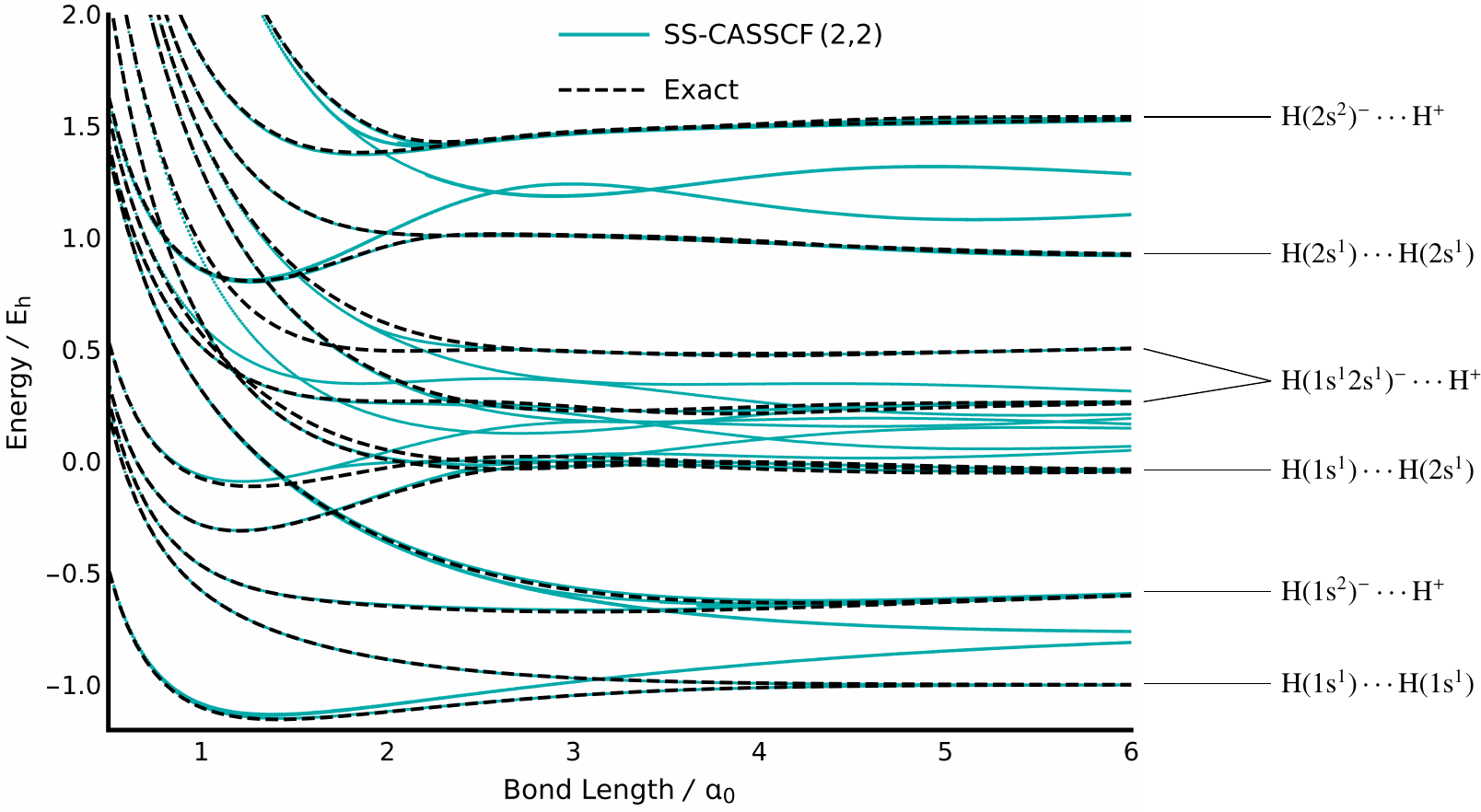}
\caption{State-specific CASSCF\,(2,2) stationary points can be identified for every 
excited FCI state in \ce{H2}. 
Additional solutions can also be found that dissociate to an unphysical electronic state.}
\label{fig:h2_binding_curve}
\end{figure*}

We start by considering the \ce{H2} binding curve using the 6-31G basis set.\cite{Ditchfield1971}
To identify all the CASSCF\,(2,2) solutions, a comprehensive search was performed using up to 1000 
random starting points for target Hessian indices from 0 to 16.
Solutions were identified near the equilibrium geometry $R=\SI{1.0}{\bohr}$ and the dissociation 
limit $R=\SI{6.0}{\bohr}$, and were then traced over all bond lengths, as shown in Fig.~\ref{fig:h2_binding_curve}.
We believe that we have found every stationary point on the landscape, although the nature
of non-convex optimisation means that this can never be guaranteed.
To the best of our knowledge, this study is the first comprehensive enumeration of the CASSCF solutions
for a molecular system.

\subsubsection{Excitations near equilibrium}

Near the equilibrium geometry, the ground state of \ce{H2} can be accurately described using a single 
reference approximation. 
We have identified 25 stationary points on the CASSCF\,(2,2) energy landscape, corresponding to
19 singlet solutions and 6 triplet solutions (Table~\ref{tab:tab_1}).
\begin{table}[htb]
  \caption{Energies of \ce{H2} at $R=1~a_0$ using the 6-31G basis set for various formalism: 
  FCI, SA-CASSCF\,(2,2), SA-CASSCF\,(3,2), and SS-CASSCF\,(2,2).}
  \label{tab:tab_1}
  \begin{ruledtabular}
    \begin{tabular}{ccccccc}
      State   &   FCI      &    SA(2,2)  &   SA(3,2)  &   SS(2,2)  & $\expval*{S^2}$ & Index\\
      \hline
            0 &   -1.09897 &    -1.07170 &   -1.08924 &   -1.09225 & 0 & 0 \\
              &            &             &            &   -1.08569 & 0 & 1 \\
              &            &             &            &   -1.07871 & 0 & 2 \\
            1 &   -0.57616 &    -0.57166 &   -0.57406 &   -0.57417 & 2 & 1 \\
            2 &   -0.46395 &    -0.43494 &   -0.44196 &   -0.46368 & 0 & 2 \\
            3 &   -0.28180 &             &   -0.27990 &   -0.27990 & 2 & 2 \\
            4 &   -0.07450 &             &   -0.06164 &   -0.05946 & 0 & 3 \\
            5 & \hd0.32015 & \hd0.33066  & \hd0.32624 & \hd0.31914 & 0 & 3 \\
              &            &             &            & \hd0.31821 & 0 & 2 \\
              &            &             &            & \hd0.31844 & 0 & 2 \\
              &            &             &            & \hd0.32440 & 0 & 3 \\
            6 & \hd0.51519 &             & \hd0.51654 & \hd0.51638 & 2 & 3 \\
            7 & \hd0.57224 &             & \hd0.61682 & \hd0.61429 & 0 & 4 \\
            8 & \hd0.62520 &             &            & \hd0.62401 & 2 & 3 \\
            9 & \hd0.86353 &             & \hd0.86876 & \hd0.86392 & 0 & 5 \\
              &            &             &            & \hd0.85673 & 0 & 4 \\
              &            &             &            & \hd0.86266 & 0 & 4 \\
            10 & \hd0.96373 &             &            & \hd0.91147 & 0 & 4 \\
           11 & \hd1.30761 &             &            & \hd1.30572 & 2 & 4 \\
           12 & \hd1.46479 &             &            & \hd1.45704 & 0 & 5 \\
           13 & \hd1.61884 &             &            & \hd1.61685 & 2 & 5 \\
           14 & \hd1.81277 &             &            & \hd1.80747 & 0 & 6 \\
           15 & \hd2.71948 &             &            & \hd2.71766 & 0 & 7 \\
              &            &             &            & \hd2.70046 & 0 & 6 \\
              &            &             &            & \hd2.69883 & 0 & 5 \\
    \end{tabular}
  \end{ruledtabular}
\end{table}
Each of the exact FCI states has a corresponding SS-CASSCF\,(2,2) counterpart, and the 
energetic agreement between these solutions is consistent for all excitations.
We have also found several additional solutions that appear to be less accurate 
approximations to the exact states, which will be characterised in Sections~\ref{subsubsec:closedshell}
and \ref{subsubsec:openshell}.
In comparison, the SA-CASSCF\,(2,2) approach can only describe the lowest triplet and the three
lowest singlet states, while increasing the number of active orbitals to a (3,2) active space 
provides an approximation to the lowest nine excitations.

These results demonstrate two important features of state-specific calculations.
Firstly, they can describe more excited states than state-averaged calculations by defining the 
active space using only orbitals that are relevant for a particular excitation.
This property allows higher energy excitations to be predicted while avoiding
large active spaces and the associated increase in the configuration space.
An upper bound to the exact excited state energy is only provided by 
stationary points that correspond to the correct excitation within the CASCI configuration
space,\cite{Golab1983}
although more accurate energies are generally preferred even if they are not variational. 
Secondly, bespoke orbital optimisation for each state-specific solutions can give more 
accurate total energies for the  excited states compared to the state-averaged approach. 
For example, the mean absolute deviations (MAD) for the lowest four states are $2.5\,\mathrm{mE_h}$ 
and $17.8\,\mathrm{mE_h}$ for the state-specific and state-averaged CASSCF\,(2,2) approaches, respectively.

Using analytic second derivatives of the energy also allows the nature of 
SS-CASSCF\,(2,2) stationary points to be characterised according to their number of 
downhill directions. 
The corresponding Hessian index for each solution is listed in Table~\ref{tab:tab_1}.
It is known that the exact $n$-th excited state should have $n$ downhill directions.\cite{Olsen1983,Olsen1982,Burton2022a}
We find that the SS-CASSCF\,(2,2) excited states are all saddle points on the electronic energy 
landscape and the Hessian index generally increases with the energy,
in common with the observations for other theoretical approximations.\cite{Gilbert2008,Hait2021,Burton2021a,Kossoski2021}
However, except for the lowest three exact states, the Hessian index does not provide a reliable 
indicator of the corresponding exact excitation index.
This mismatch must always occur for higher-lying excited states as the 
approximate CASSCF\,(2,2) wave function has fewer degrees of freedom than the exact formulation.
Consequently, if we only consider stationary points of the correct Hessian index, then 
we must forgo the advantages of capturing state-specific excitations outside 
the state-averaged active space.


\subsubsection{Multiple ground state solutions}
\label{subsubsec:closedshell}


While Table~\ref{tab:tab_1} shows that a SS-CASSCF\,(2,2) approximation can be identified
for each exact eigenstate, we also find additional state-specific solutions.
In particular, there are three close-lying stationary points that can be considered as 
approximations to the ground state, with Hessian indices of 0, 1, and 2 in order of
ascending energy.
This pattern of multiple solutions is repeated for the $(2\sigg)^2$ and $(2\sigu)^2$ 
singlet configurations, while the other closed-shell $(1\sigu)^2$
configuration exhibits four close-lying solutions.
Choosing the most physical solution for each eigenstate presents a challenge
for state-specific CASSCF approaches.
Therefore, it is important that we understand their mathematical origins and physical differences.

\begin{figure*}
\includegraphics[width=0.85\linewidth]{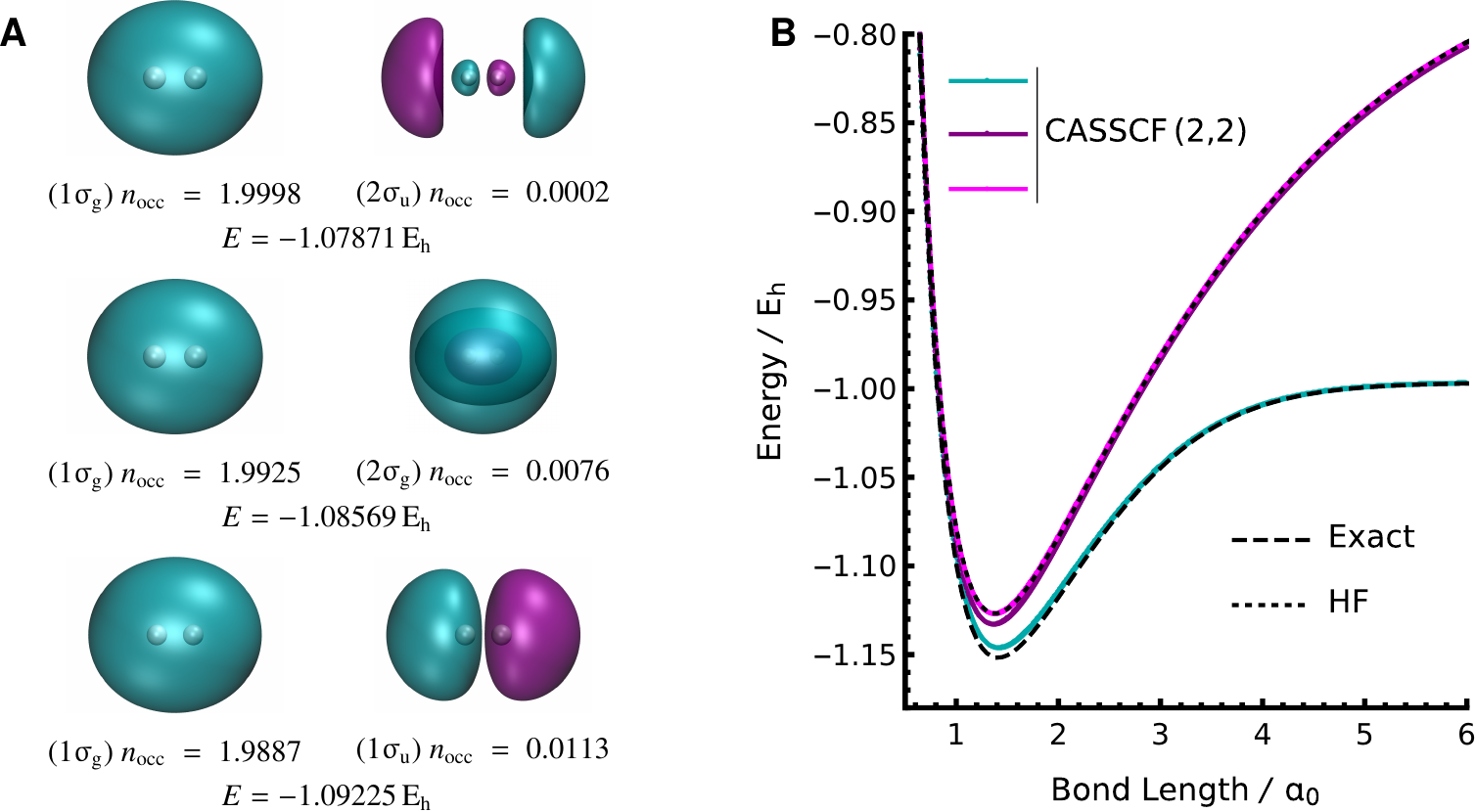}
\caption{There are three SS-CASSCF(2,2) solutions that represent the exact ground state in \ce{H2}.
(\textsf{\textbf{A}}) Comparison of the natural orbitals for each ground-state solution at $R=\SI{1.0}{\bohr}$. 
(\textsf{\textbf{B}}) Only the lowest-energy solution dissociates correctly, while the higher-energy 
solutions mirror the restricted Hartree--Fock binding curve.}
\label{fig:multigs_h2}
\end{figure*}

The natural orbitals in the active space provides a clear explanation for the multiple \ce{H2} ground-state solutions.
Figure~\ref{fig:multigs_h2}\textcolor{blue}{A} compares the natural orbitals and occupation numbers for the three
lowest-energy singlet stationary points.
Since the ground state at the equilibrium geometry can be relatively well approximated by 
a single closed-shell Slater determinant, the active space for
each of these solutions includes a $(1\sigg)$-like natural orbital that is almost completely 
doubly occupied.
This natural orbital dominates the electronic wave function and the corresponding energies 
are all relatively close approximations to the exact ground state.
However, the second active orbital, which is almost completely unoccupied, is different for each
solution, corresponding to a $(1\sigu)$, $(2\sigg)$, or $(2\sigu)$ orbital as the energy increases,
respectively.
These higher-energy stationary points have downhill orbital rotations that interconvert the 
multiple ground state solutions and correspond to the negative eigenvalues of the Hessian.

Different choices for the nearly unoccupied active orbital have only a small effect on 
the total \ce{H2} energy near equilibrium.
However, the incorrect choice of the active space becomes very significant as 
the bond is stretched towards dissociation.
Only the $\{1\sigg,1\sigu\}$ active space can correctly dissociate into the $\mathrm{H(1s)\cdots H(1s)}$
ground state of the dissociated fragments (Fig.~\ref{fig:multigs_h2}\textcolor{blue}{B}).
In contrast, the binding curves for the $\{ 1\sigg,2\sigg \}$ and $\{ 1\sigg,2\sigu \}$ solutions
mirrors the RHF energy as the corresponding wave functions are close to a single Slater 
determinant at all geometries, with $(1\sigg)$ occupation numbers at dissociation of 1.997 and 1.999, respectively.
Notably, the stationary points preserve the character of the active orbitals along the potential energy 
surface, suggesting that SS-CASSCF solutions exhibit some degree of diabatic character.

The same pattern of solutions is observed for the other closed-shell solutions. 
However, the $(1\sigu)^2$ configuration exhibits an additional multiple solution 
where the nearly unoccupied active orbital corresponds to a symmetry-broken 
$2\mathrm{s}$-like orbital localised on either 
the left or right \ce{H} atom.
This symmetry breaking results in a two-fold degenerate pair of stationary points.

These results indicate that additional solutions can arise from the free choice of virtual orbitals 
when the active space is larger than required for the degree of static correlation.
Malrieu and co-works elegantly summarised this phenomenon by stating that
\textit{``the so-called valence CASSCF wave function does not necessarily keep a valence character when the wave function concentrates
on a closed-shell valence bond structure''.}\cite{Meras1990}
Therefore, we expect that the number of ground state solutions will increase combinatorially 
with the number of active orbitals or the basis set size, 
and the number of unphysical solutions can grow for larger active spaces even though 
the correct ground state solution will become more accurate.
Table~\ref{tab:H2_6_311g} demonstrates this increase for \ce{H2} using the 6-311G basis set
with three basis functions for each hydrogen atom.\cite{Krishnan1980}
Taking the (3,2) active space as an example, there are two redundant active orbitals beyond the 
$1\sigg$ that must be chosen from the five remaining orbitals, giving a total of ten solutions through 
the binomial coefficient $\binom{5}{2}=10$.
The relative energy ordering of these additional solutions will depend on the amount of dynamic 
correlation captured by the redundant active orbitals, which may not correspond with the same 
orbital required to capture the static correlation in the dissociation limit.
This phenomenon has previously been described for \ce{MgO}, where oxygen-centred orbitals are preferred over
the magnesium d orbitals,\cite{Hanscam2022} and transition metal compounds where 
non-valence $\mathrm{d}$ orbitals may be preferred over certain  valence d orbitals.\cite{Andersson1992b}

\begin{table}[htb]
  \caption{Close-lying ground-state $(n,2)$ SS-CASSCF energies ($\Eh$) of \ce{H2} at $R=1~a_0$ 
   using the 6-311G basis set for various active space size $n$.}
    \label{tab:H2_6_311g}
  \begin{ruledtabular}
    \begin{tabular}{llllll}
      SS(1,2): HF & -1.08025 & & & &\\
      \hline
      SS(2,2) & -1.09429 & -1.08866 & -1.08074 & -1.08033 & -1.08026 \\
      \hline
      SS(3,2) & -1.10195 & -1.09500 & -1.09436 & -1.09429     & -1.08904 \\
                   & -1.08886 & -1.08867 & -1.08082 & -1.08075 & -1.08034  \\
      \hline
      SS(4,2) & -1.10251 & -1.10212 & -1.10196 & -1.09507 & -1.09500 \\
                   & -1.09437  & -1.08923 & -1.08905 & -1.08886 & -1.08083  \\
      \hline
      SS(5,2) & -1.10267 & -1.10251 & -1.10213 & -1.09507 & -1.08924 \\
      \hline
      SS(6,2): FCI & -1.10267 & & & &\\
    \end{tabular}
  \end{ruledtabular}
\end{table}
\subsubsection{Open-shell singlet and triplet excitations}
\label{subsubsec:openshell}

\begin{figure}
  \centering
  \includegraphics[width=\linewidth]{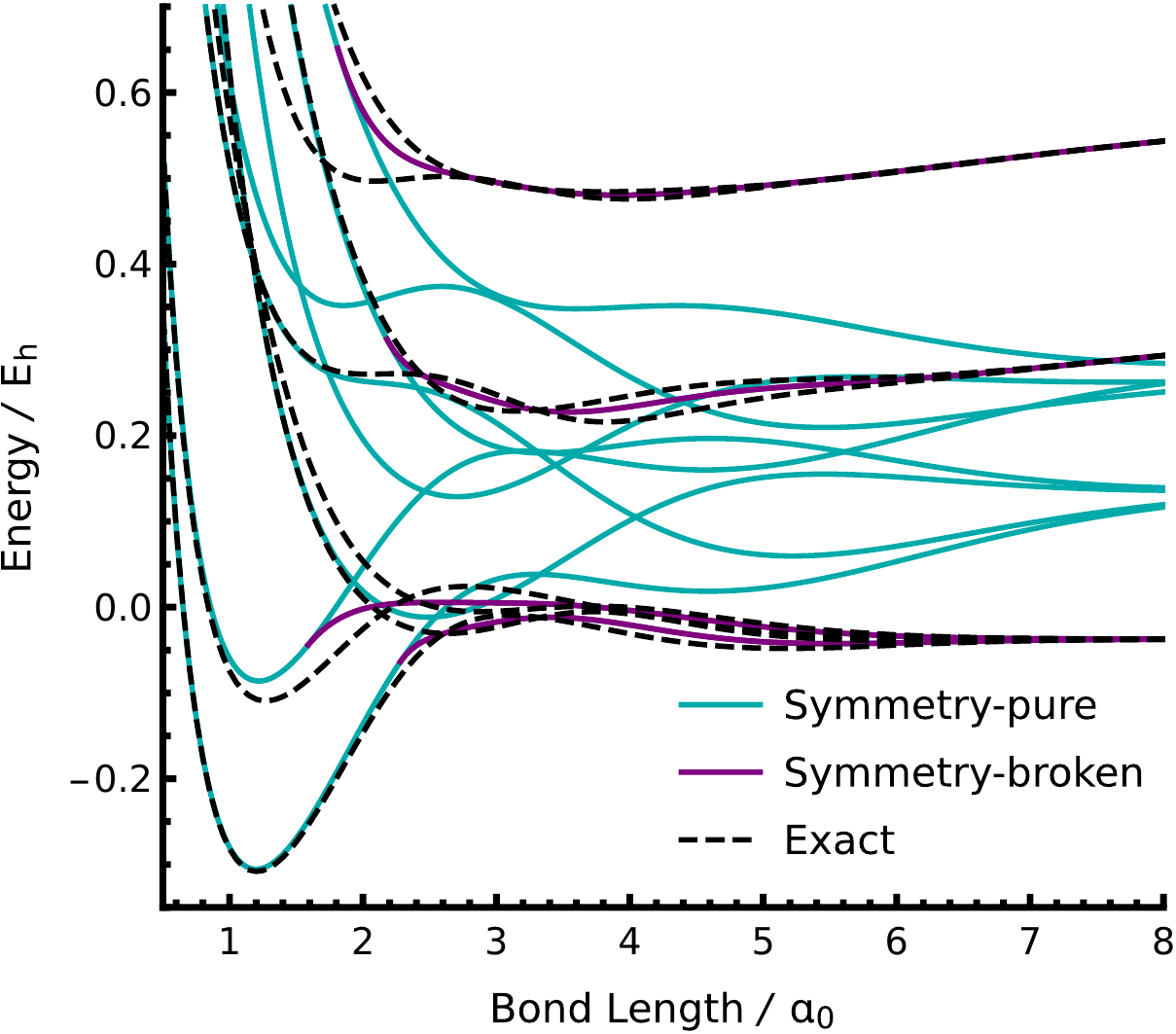}
  \caption{Spontaneous symmetry breaking occurs when the active space is not large enough to capture 
  all the important configurations in the physical wave function, as illustrated for the $1\mathrm{s}\,2\mathrm{s}$ 
  states in the dissociation of \ce{H2} (6-31G).}
  \label{fig:h2_symmetry_breaking}
\end{figure}

The low-lying open-shell triplet and singlet $(1\sigg)^{1} (1\sigu)^{1}$ configurations
are represented by only one SS-CASSCF\,(2,2) solution across the full 
binding curve (Fig.~\ref{fig:h2_binding_curve}).
These single solutions arise because all the active orbitals are required to describe the two-configurational
static correlation and there is no flexibility for multiple solutions to exist.
In addition, SS-CASSCF\,(2,2) gives an accurate representation of the open-shell 
$(1\upsigma_\textrm{g/u})^{1} (2 \upsigma_\textrm{g/u})^{1}$ configurations.
However, the accuracy of these solutions deteriorates in the dissociation limit, where additional 
symmetry broken solutions can be identified (Fig.~\ref{fig:h2_symmetry_breaking}).
These additional solutions break spatial symmetry and spontaneously 
appear at instability thresholds that are multi-configurational analogues to the Coulson--Fischer 
points\cite{Coulson1949} in HF theory.\cite{Fukutome1973,Fukutome1975,Mestechkin1978,Mestechkin1979,Mestechkin1988}
Each stationary point is a pure singlet or triplet state and has a two-fold degeneracy, 
reflecting the left-right symmetry of the molecule.

The origin of this symmetry breaking is explained by considering the correlation processes
involved in the excited dissociation limit.
These excited states dissociate to hydrogenic $(1\mathrm{s})^1 (2\mathrm{s})^1$ configurations, where the occupied 
$1\mathrm{s}$ and $2\mathrm{s}$ orbitals can either be on the same or different atomic centres.
Taking the latter case as an example, the corresponding open-shell singlet wave function
at large nuclear separations has the form
\begin{equation}
\begin{split}
\ket{\Psi} &= \frac{1}{2}
\Big(\ket{1\s_\text{L}2\s_\text{R}} + \ket{1\s_\text{R}2\s_\text{L}}\Big)
\Big(\ket{\alpha\beta} - \ket{\beta \alpha}\Big).
\end{split}
\label{eq:opensinglet}
\end{equation}
Correctly describing this wave function requires an active space with four spatial orbitals
$\{1\s_\text{L}, 1\s_\text{R},2\s_\text{L}, 2\s_\text{R}  \}$, or equivalently $\{1\sigg, 1\sigu, 2\sigg, 2\sigu\}$,
and thus the SS-CASSCF\,(2,2) approximation is insufficient for these correlation mechanisms.
Instead, the symmetry breaking reduces the SS-CASSCF\,(2,2) wave function to a subset of the dominant configurations, 
e.g.\
\begin{equation}
\ket*{\tilde{\Psi}} = \frac{1}{\sqrt{2}} \ket{1\s_\text{L}2\s_\text{R}}
\Big(\ket{\alpha\beta} - \ket{\beta \alpha} \Big).
\end{equation}
The CASSCF configurations corresponding to each symmetry-broken solution are assigned in Table~\ref{tab:h2_sym_breaking}.
\begin{table}[h]
\begin{ruledtabular}
\begin{tabular}{cccc}
State & Energy / $\Eh$ & $\expval*{S^2}$ & Configuration \\ \hline
A & \hphantom{-}0.543355 & 0.00 
& $\begin{cases}%
\ket{1\s_\text{L}2\s_\text{L}} \qty(\ket{\alpha\beta} - \ket{\beta \alpha} ) \\ 
\ket{1\s_\text{R}2\s_\text{R}} \qty(\ket{\alpha\beta} - \ket{\beta \alpha} ) 
\end{cases}$ 
\\ \hline
B & \hphantom{-}0.293363 & 2.00 
& $\begin{cases}%
\ket{1\s_\text{L}2\s_\text{L}} \qty(\ket{\alpha\beta} + \ket{\beta \alpha} ) \\ 
\ket{1\s_\text{R}2\s_\text{R}} \qty(\ket{\alpha\beta} + \ket{\beta \alpha} ) 
\end{cases}$ 
\\ \hline
C & -0.037221 & 0.00
& $\begin{cases}%
\ket{1\s_\text{L}2\s_\text{R}} \qty(\ket{\alpha\beta} - \ket{\beta \alpha} ) \\ 
\ket{1\s_\text{R}2\s_\text{L}} \qty(\ket{\alpha\beta} - \ket{\beta \alpha} ) 
\end{cases}$ 
\\ \hline
D & -0.037499 & 2.00 
& $\begin{cases}%
\ket{1\s_\text{L}2\s_\text{R}} \qty(\ket{\alpha\beta} + \ket{\beta \alpha} ) \\ 
\ket{1\s_\text{R}2\s_\text{L}} \qty(\ket{\alpha\beta} + \ket{\beta \alpha} ) 
\end{cases}$ 
\end{tabular}
\end{ruledtabular}
\caption{The symmetry-broken CASSCF\,(2,2) solutions in the dissociation of \ce{H2} 
are two-fold degenerate and represent dominant configurations in the 
exact excitations.}
\label{tab:h2_sym_breaking}
\end{table}
This ``pinning'' of the wave function onto a particular electronic configuration is directly analogous to the 
symmetry breaking phenomena observed in HF theory\cite{Trail2003,Burton2021b}
and demonstrates that the active space is too small to fully account for the static correlation.

From the energy landscape perspective, the onset of symmetry-broken CASSCF\,(2,2) states
is associated with a change in the Hessian index for the associated symmetry-pure solutions.
For example, the symmetry-broken state D (Table~\ref{tab:h2_sym_breaking}) emerges from the symmetry-pure 
$(1\upsigma_\mathrm{g})^1 (2\upsigma_\mathrm{g})^1$ 
triplet state at an instability threshold close to $R = \SI{2.28}{\bohr}$.
The Hessian index of the symmetry-pure state changes from 2 to 3 at this point, while the 
symmetry-broken solutions form index-2 saddle points, leading to a higher-index analogue of a 
cusp catastrophe.\cite{GilmoreBook,Burton2018,Burton2021a}
Practically, the emergence of a zero Hessian eigenvalue at these instability thresholds may hinder 
the numerical optimisation of second-order techniques onto these higher-energy stationary points.
It is also interesting to note that, 
while the symmetry-broken solutions describe two degenerate FCI states at dissociation,
they only connect to one of the corresponding symmetry-pure solutions in the equilibrium region.
Consequently, one cannot rely on these additional solutions to obtain an accurate and  
continuous representation of every excited state across all geometries.

\subsection{Conical Intersection in methylene}

We next consider the bending mode of methylene, which has a diradical ground state with $^3\mathrm{B}_1$ 
symmetry and a low-lying $1\,^1\mathrm{A}_1$ excited state. 
The bond length was fixed to the value $R(\ce{C-H}) = \SI{2.11}{\bohr}$ identified by 
Bauschlicher and Taylor\cite{Bauschlicher1986a,Bauschlicher1986b} 
and the 6-31G basis set was used.\cite{Ditchfield1971}
Methylene has a long history as a benchmark for electronic structure theory.\cite{Schaefer1986}
One of the primary questions is the description of the conical intersection between
 the low-lying $^3\mathrm{B}_1$ and $1\,^{1}\mathrm{A}_1$ states.


\begin{figure*}[htb]
\includegraphics[width=\linewidth]{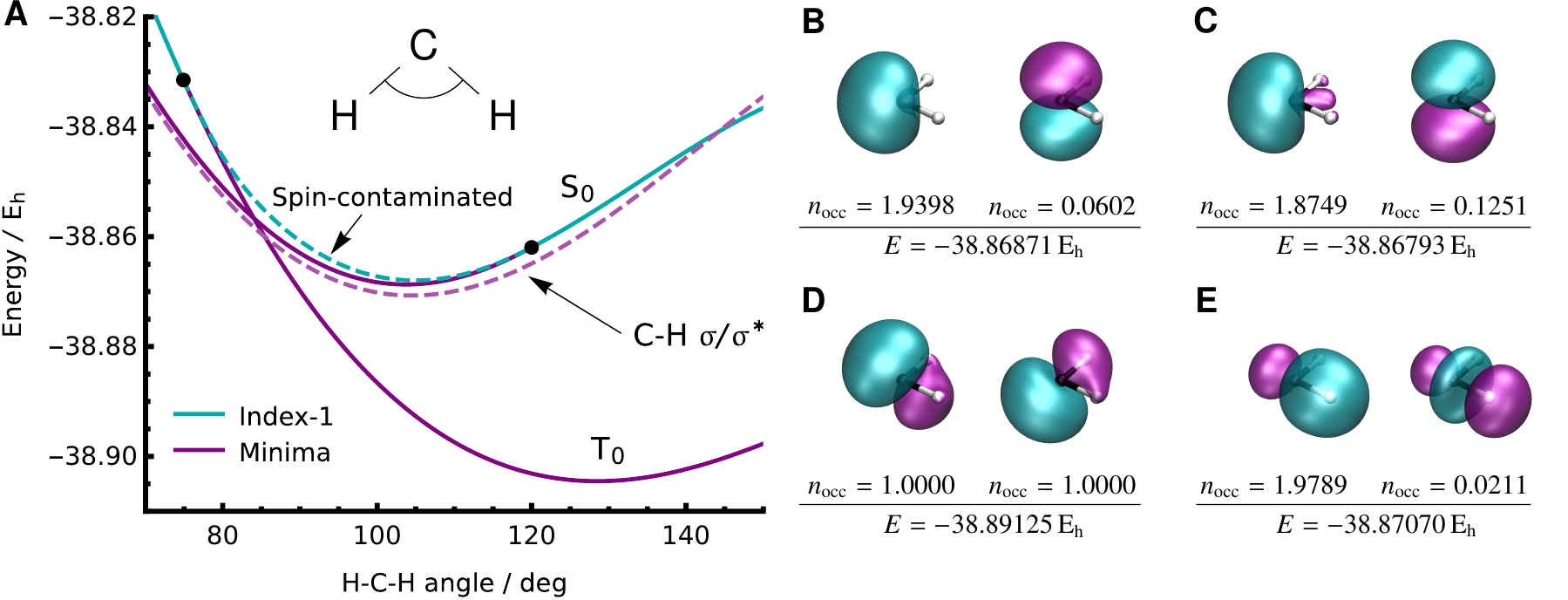}
\caption{Low-lying SS-CASSCF\,(2,2) in the bending mode of methylene represent the 
$1\,^3\mathrm{B}_1$ and $1\,^1\mathrm{A}_1$ configurations.
(\textbf{\textsf{A}}) Both states remain local minima (solid purple) for a short region beyond the conical intersection before 
becoming an index-1 saddle point (solid cyan). 
An additional spin-contaminated index-1 saddle point (dashed cyan) connects the two instability thresholds (black dots).
Two degenerate local minima exist everywhere along the bending curve (dashed purple) with an active space containing
C-H bonding $\upsigma$ and antibonding $\upsigma^{*}$ orbitals.
(\textbf{\textsf{B}}--\textbf{\textsf{E}}) The natural orbitals at a bond angle of $103.7^{\circ}$ are illustrated 
for each solution.}
\label{fig:methylene_bend}
\end{figure*}

\subsubsection{Local minima for the minimal (2,2) active space}
A minimal two-configuration wave function is required to qualitatively describe 
both the lowest-energy singlet $\tS_0$ ($^1\mathrm{A}_1$)
and diradical triplet $\tT_0$ ($^3\mathrm{B}_1$) states.\cite{Bauschlicher1986a}
Therefore, we begin by analysing the SS-CASSCF\,(2,2) energy landscape.
The $\tS_0$ and $\tT_0$ are the ground state for small and large bond angles, respectively, 
and provide an example of a conical intersection separating the two regimes. 
At bond angles of $76^\circ$, $102^\circ$ and $130^\circ$, a large number of stationary points can be identified
with a variety of Hessian indices.
Therefore, we simplify our analysis by focussing on a subset of low-energy solutions that
resemble the desired physical states (Fig.~\ref{fig:methylene_bend}).

The energetic minimum of the $\tS_0$ state occurs at a bond angle of $103.7^\circ$.
While the $\tS_0$ state is the first excited state at this geometry, we find that 
the corresponding SS-CASSCF\,(2,2) stationary point is a local minimum rather than an index-1 saddle point.
This incorrect Hessian index arises from a root flip in the configuration space, 
where the singlet state is the ground state for the corresponding active orbitals. 
When the bond angle increases, this singlet state eventually becomes an index-1 saddle point.
Similarly, when the bond angle decreases from $103.7^\circ$, 
the $\tT_0$ state remains a local minimum beyond the conical intersection where it becomes the first excited state.
This process behaves like an unphysical hysteresis, where the ground state remains a local minimum for a
small region after a conical intersection before becoming an index-1 saddle point at an instability threshold.

An additional index-1 saddle point can be identified that connects these two solutions and coalesces with 
each local minimum at the two instability thresholds.
This unphysical index-1 stationary point is two-fold degenerate,
has symmetry-pure spatial orbitals, but is spin contaminated with
an $\expval*{\hat{S}^2}$ value that changes continuously from 0 to 2 as it connects the singlet and 
triplet states.
Similar patterns of coalescing solutions have been observed in single determinant 
SCF approximations,\cite{Burton2018,Zarotiadis2020,Fukutome1975,Huynh2019}
particularly in the generalised HF representation of conical intersections between 
states with different $\expval*{\hat{S}_z}$ values.\cite{Jimenez-Hoyos2011}
In contrast to symmetry-broken SCF solutions, the spin contamination observed here arises from the 
mixture of singlet and triplet states in the configurational part of the CASSCF wave function.

Since the $\tS_0$ solution has only one significantly occupied active orbital,
we predict the existence of closely-related solutions that have alternative redundant orbitals 
with $n_\text{occ} \approx 0$.
Indeed, there are a pair of degenerate local minima that lie slightly lower in energy 
than the $\tS_0$ solution. 
In contrast to the \ce{H2} ground state, including the inactive space means that methylene has multiple doubly occupied orbitals, 
and thus the active orbital with $n_\text{occ} \approx 2$ may also change between different solutions.
The active orbitals for these symmetry-broken solutions are localised bonding $\upsigma$ and 
anti-bonding $\upsigma^*$ orbitals for one of the two \ce{C-H} bonds, and the degeneracy accounts for the
two possible ways to localise onto one bond.
Notably, the symmetry breaking here is associated with an active space that is too large, in contrast
to \ce{H2} where symmetry breaking arises from an insufficient active space for the static correlation. 
These solutions are local minima across all the bond angles considered.
While they provide an accurate energy for the $\tS_0$ state near the singlet equilibrium structure, this
deteriorates for large angles as the active space cannot describe the diradical 
open-shell $^1\Sigma_\text{g}^{+}$ state at the linear geometry.
Their existence indicates that the \ce{C-H} $\upsigma/\upsigma^{*}$ configurations provide 
an important contribution to static correlation and should ideally be included in the active space, 
as suggested by Bauschlicher and Taylor.\cite{Bauschlicher1986a,Bauschlicher1986b} 

\subsubsection{Full valence active space}
\begin{figure}[htb]
\includegraphics[width=\linewidth]{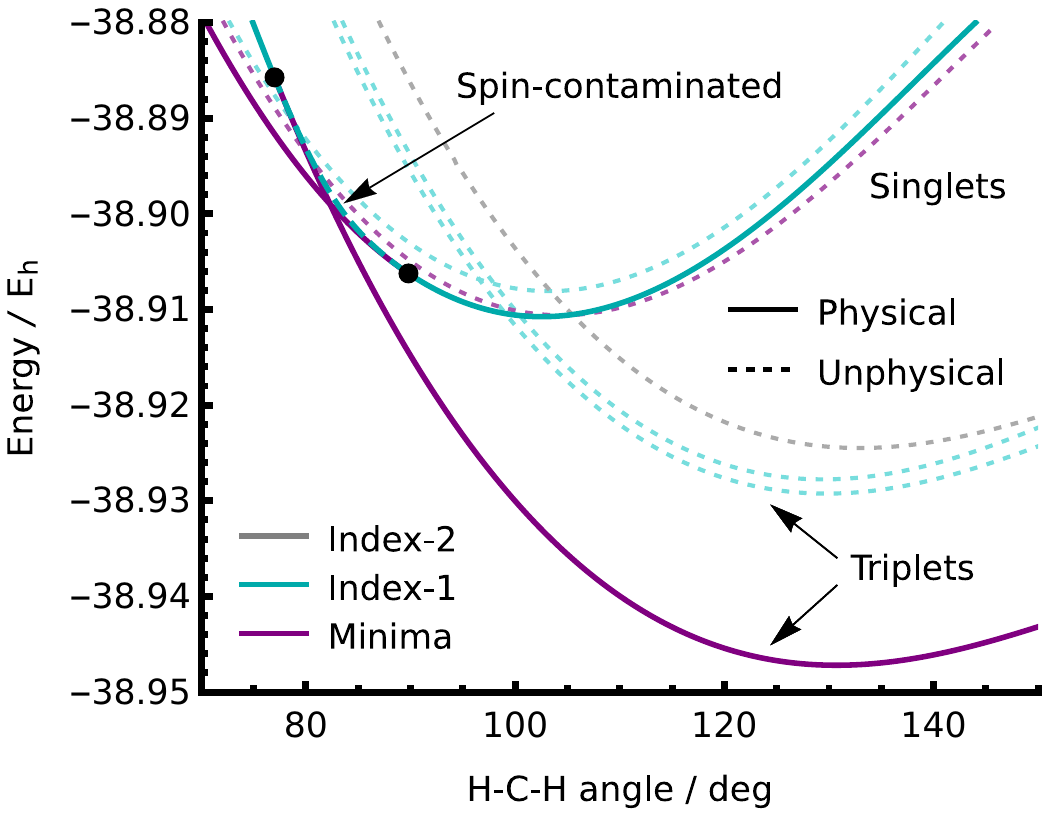}
\caption{Low-lying SS-CASSCF\,(6,6) for the bending mode of methylene 
represent the $1\,^3\mathrm{B}_1$ and $1\,^1\mathrm{A}_1$ configurations.
The full valence (6,6) active space introduces more unphysical solutions, but does not
remove the spin-contaminated solution that arises at the conical intersection.}
\label{fig:methylene_6_6}
\end{figure}

\begin{figure*}[htb]
\includegraphics[width=\linewidth]{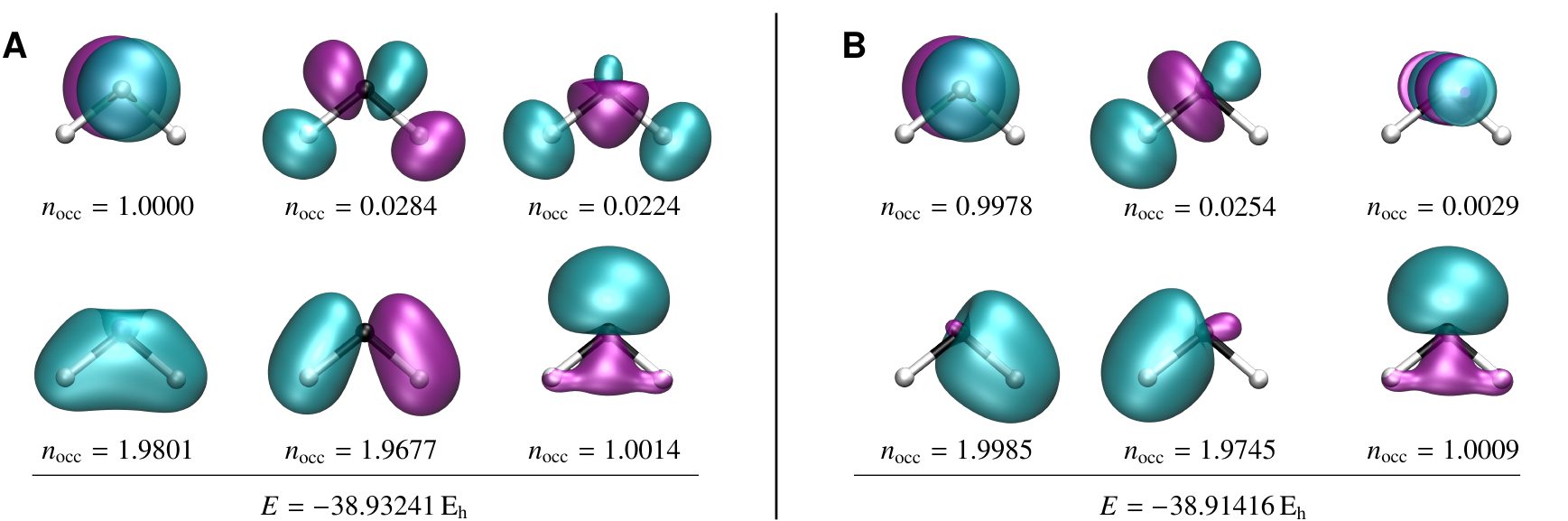}
\caption{Comparison of the active orbitals for the two lowest energy triplet CASSCF solutions for \ce{CH2} (6-31G) 
using the full valence (6,6) active space at a bond angle of $102^\circ$. 
(\textbf{A}) The active orbitals for the local minimum represent the chemically intuitive valence space.
(\textbf{B}) For the unphysical index-1 saddle point, one of the antibonding \ce{C-H} $\upsigma^*$
orbitals with $n_\text{occ} \approx 0$ is replaced by a carbon $3 \mathrm{p}$ orbital with $n_\text{occ} = 0.0029$.
The remaining $\upsigma$ and $\upsigma^{*}$ orbitals localise onto the \ce{C-H} bonds.}
\label{fig:ch2_6-6_orbitals}
\end{figure*}

Using the full-valence (6,6) active space, we find that the symmetry-pure singlet state is now
correctly represented by an index-1 saddle point at a bond angle of $\SI{102}{\degree}$ 
(Fig.~\ref{fig:methylene_6_6}).
The unique downhill direction corresponds to a rotation in the configuration space only, as expected for 
the first excited state.
Despite the larger active space, a root flip still occurs as the states approach the singlet--triplet 
conical intersection at $\SI{82.2}{\degree}$, with the singlet state becoming a local minimum 
at $\SI{89.6}{\degree}$ and the triplet state becoming an index-1 saddle point at $\SI{77.0}{\degree}$.
Like the (2,2) active space, a degenerate pair of unphysical, spin-contaminated index-1 saddle 
points connect the solutions that cross at the conical intersection.
This phenomenon occurs because the orbital optimisation can lower the energy of the target excited state 
below the corresponding ground state configuration when the energy gap becomes small.
Therefore, while larger active spaces will reduce the range of molecular 
geometries affected, these
unphysical local minima will be common for state-specific conical intersections.

While the larger active space  alleviates root flipping, it also causes 
more unphysical solutions associated with redundant active orbitals.
For example, the triplet ground state (dominated by two configurations) is represented
by one  SS-CASSCF\,(2,2) solution, but there are several higher-energy solutions in 
the (6,6) active space.
Analogously to the \ce{H2} ground state, these additional solutions have a higher Hessian index, 
with two index-1 and one index-2 saddle points represented in Fig.~\ref{fig:methylene_6_6}.
Again, the main difference from the true ground state are the
active orbitals with occupation numbers close to zero, as illustrated
for the global minimum and lowest-energy index-1 saddle point in Fig.~\ref{fig:ch2_6-6_orbitals}.
Furthermore, we find an additional local minimum and index-1 saddle point that represent the
$^1\mathrm{A_1}$ state.
While all the triplet solutions give approximately the same equilibrium bond angle, 
the unphysical stationary points shift the conical intersection to coincide with the 
singlet equilibrium geometry.
This qualitative change in the energy surface would create a near-barrierless decay from the 
singlet excited state to the triplet ground state, demonstrating the importance of verifying the physicality of 
state-specific solutions.

\subsection{Avoided crossing in Lithium Fluoride}
\label{sec:avoided}

\subsubsection{Physicality of multiple solutions}

\begin{figure*}[htb]
\includegraphics[width=\linewidth]{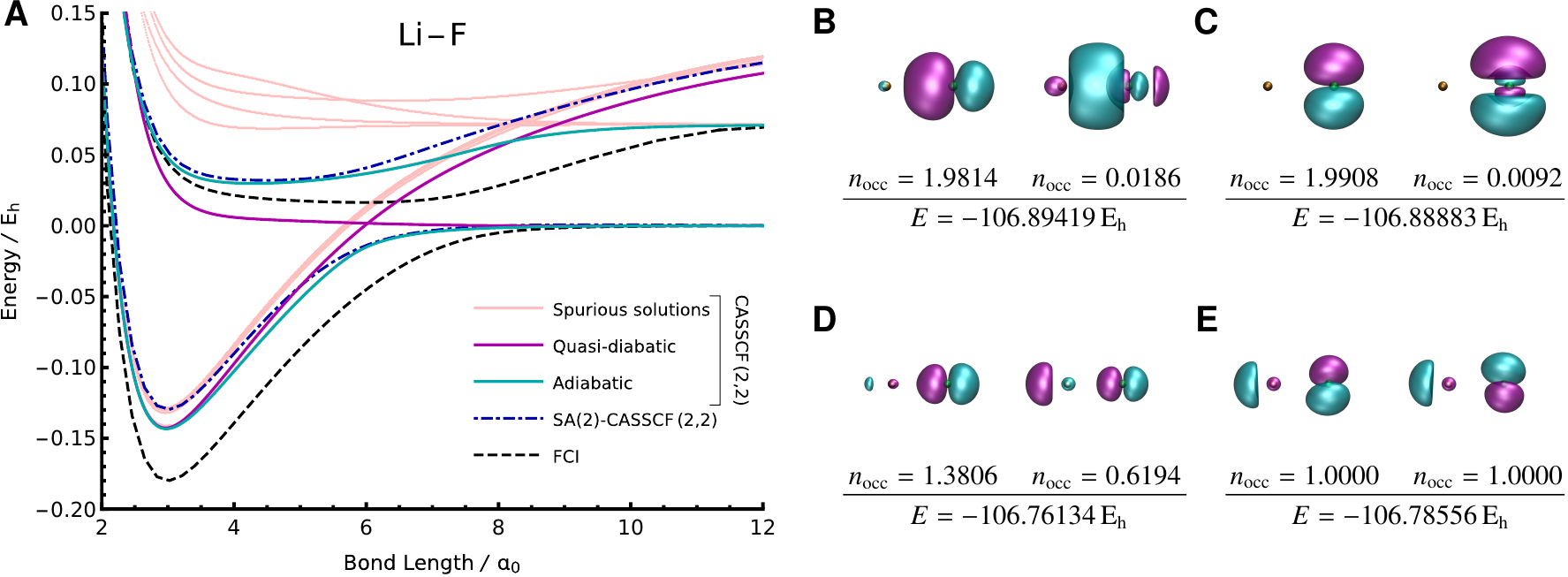}
  \caption{(\textbf{A}) The SS-CASSCF\,(2,2) appoach gives many solutions for the \ce{LiF} binding curve (6-31G) 
  when the ground or excited state is dominated by a single configuration.
  Ground- and excited-state solutions with a suitable active space (\textbf{B} and \textbf{D}) behave adiabatically 
  at the avoided crossing (cyan lines).
  Additional solutions with unsuitable active orbitals can represent either the 
  ionic equilibrium configuration (\textbf{C}) or the covalent dissociation configuration (\textbf{E}),
  and behave quasi-diabatically at the avoided crossing (purple lines).
  The active orbitals are plotted at $R(\ce{Li-F})=\SI{4}{\bohr}$.
  Exact FCI and SA(2)-CASSCF\,(2,2) data are taken from Ref.~\onlinecite{Burton2020}.}
\label{fig:LiF_6-31g}
\end{figure*}

The \ce{LiF} binding curve provides a typical example of an avoided crossing.
The ground state has ionic character at equilibrium, but becomes a covalent state with almost no dipole moment
in the dissociation limit.
Multiple HF solutions are known to behave ``quasi-diabatically'' and cross each other at the 
physical avoided crossing.\cite{Thom2009,Burton2020}
On the other hand, Bauschlicher and Langhoff demonstrated that this avoided crossing can lead to 
discontinuities in the CASSCF ground- and excited-state energy surfaces.\cite{Bauschlicher1988}
Here, we start by considering the state-specific singlet CASSCF solutions in the 6-31G basis set.

Using the minimal (2,2) active, we search for stationary points with 
Hessian indices of 0 to 10 at $R(\ce{Li-F}) = \SI{2.75}{\bohr}$ (near the equilibrium geometry), using 
1000 random starting points for each index.
The active space for the SS-CASSCF global minimum contains the valence bonding $\upsigma$ and anti-bonding $\upsigma^{*}$
orbitals with occupation numbers close to 2 and 0, respectively (Fig.~\ref{fig:LiF_6-31g}\textcolor{blue}{B}).
Because the exact wave function is dominated by a single closed-shell configuration, 
there are many additional solutions that are close to the ground-state energy at the equilibrium geometry.
For example, the second lowest energy solution has an active space containing the out-of-plane 
fluorine $2\mathrm{p_{x/y}}$ and $3\mathrm{p_{x/y}}$ orbitals with occupation numbers close to 2 and 0, 
respectively (Fig.~\ref{fig:LiF_6-31g}\textcolor{blue}{C}).
This active space accounts for the radial correlation on the fluorine atom, providing a more balanced 
description of \ce{F} and \ce{F-}.\cite{Bauschlicher1988}
In contrast, the exact excited state is more multiconfigurational at short bond lengths and is accurately represented 
by only one solution  (Fig.~\ref{fig:LiF_6-31g}\textcolor{blue}{D}), alongside a spurious symmetry-broken 
solution with diradical character (Fig.~\ref{fig:LiF_6-31g}\textcolor{blue}{E}).
These characteristics are reversed for bond lengths longer than the avoided crossing, 
where the excited state has closed-shell character with a large number of solutions
and the ground state is represented by only two solutions.

State-specific CASSCF solutions can behave both quasi-diabatically and adiabatically in the vicinity 
of the avoided crossing. 
As the bond length changes, the unphysical solutions do not have the correct active orbitals
to capture the strong correlation at the avoided crossing.
Therefore, the two lowest-energy unphysical solutions intersect quasi-diabatically 
(dark purple in Fig.~\ref{fig:LiF_6-31g}\textcolor{blue}{A}, corresponding to the solutions 
in Fig.~\ref{fig:LiF_6-31g}\textcolor{blue}{C} and \ref{fig:LiF_6-31g}\textcolor{blue}{E}).
On the other hand, the physically meaningful solutions behave adiabatically and correctly avoid each other 
(cyan in Fig.~\ref{fig:LiF_6-31g}\textcolor{blue}{A}).
In principle, a linear expansion of both the quasi-diabatic and adiabatic states may provide a
more accurate representation of the avoided crossing by introducing some of the dynamic correlation 
captured by the unphysical solutions. 
This expansion would require a multiconfigurational variant of nonorthogonal CI,\cite{Thom2009} where the 
Hamiltonian and overlap matrix elements can be efficiently computed using the nonorthogonal framework
developed in Refs.~\onlinecite{Burton2021c,Burton2022c}.

While a complete description of the avoided crossing requires dynamic correlation,\cite{Malrieu1995}
the advantage of state-specific orbital relaxation is still clear in the dissociation limit.
The physical SS-CASSCF excited state tends towards the exact FCI energy for the separated \ce{Li^+ \cdots F^{-}}
configuration, while state-averaged calculations (with an equal weighting for the two states)
overestimates the energy of this excited state (Fig.~\ref{fig:LiF_6-31g}\textcolor{blue}{A}).
In this SS-CASSCF solution, the $\upsigma$ and $\upsigma^{*}$ orbitals 
(Fig.~\ref{fig:LiF_6-31g}\textcolor{blue}{A}) both localise to give $\mathrm{2p_z}$ orbitals that
accurately represent the \ce{F^{-}} anion.
Consequently, as expected, the state-specific formalism provides a more accurate representation 
of this charge transfer excitation than a state-averaged approach.

\subsubsection{Elucidating the Bauschlicher--Langhoff discontinuity}
\label{sec:BLdiscont}

The seminal CASSCF investigation of \ce{LiF}, by Bauschlicher and Langhoff, 
highlighted the presence of a discontinuity in 
the ground-state dipole moment in the vicinty of the avoided crossing.\cite{Bauschlicher1988}
This discontinuity is a signature of a discontinuity in the wave function, which manifests
as a cusp in the corresponding energy surface.
This phenomenon, which we name the ``Bauschlicher--Langhoff discontinuity'', has long been 
used as key evidence for the potential issues of state-specific calculations in the vicinity 
of an avoided crossing.
Malrieu and co-workers attributed its origin to a near degeneracy between the closed-shell ionic and 
the open-shell covalent configurations, and described a lower-energy covalent state that 
emerges from a potential symmetry-breaking point as the bond length increases.\cite{Meras1990}
The framework developed here, and the advance in computing over the past 30 years, 
now allows this topological characterisation to be rigorously tested.

\begin{figure}[t]
 \centering
\includegraphics[width=\linewidth]{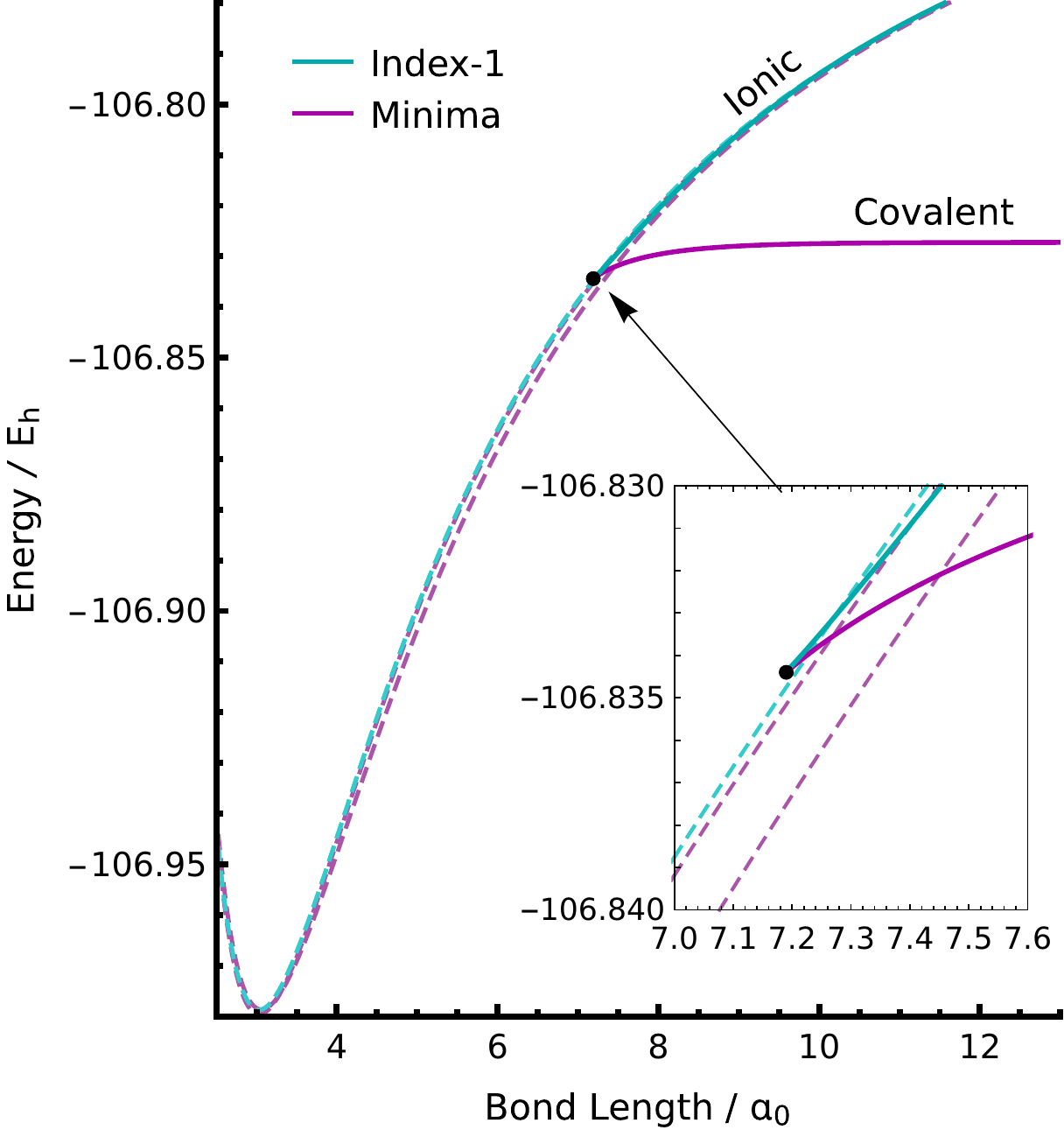}
\caption{Topology of the low-energy SS-CASSCF\,$(2,2)$ solutions near the Bausclicher--Langhoff 
discontinuity in \ce{LiF},\cite{Bauschlicher1988,Meras1990}
using the basis set defined in Ref.~\onlinecite{Bauschlicher1988}.
A cusp in the ground-state energy occurs when two local minima cross, while the covalent structure
coalesces with an index-1 saddle point at a pair annihilation point (black dot).}
\label{fig:LiF_custom}
\end{figure}

To identify the relevant solutions, we searched for minima and index-1 saddle points 
at a bond length of $\SI{8.50}{\bohr}$ using 1000 random starting points, a $(2,2)$ active space,
and the original basis set described in Ref.~\onlinecite{Bauschlicher1988}.
At $R(\ce{Li-F})=\SI{8.5}{\bohr}$, the global minimum corresponds to the covalent structure 
identified in Ref.~\onlinecite{Meras1990}.
In addition, two local minima and two index-1 saddle points exist at higher energies, 
representing the ionic configurations (Fig.~\ref{fig:LiF_custom}).
As the bond length is shortened, there is a crossing between the lowest energy ionic and covalent
minima near $R(\ce{Li-F})=\SI{7.4}{\bohr}$, which we believe corresponds to the previously 
described discontinuity.\cite{Bauschlicher1988,Meras1990}

Topologically, two non-degenerate minima cannot 
coalesce without the presence of an index-1 saddle point, and thus the disappearance described 
by Malrieu and co-workers cannot be the full picture.\cite{Meras1990}
Instead, we find that the covalent structure crosses the two lowest-energy local minima and 
eventually coalesces with an index-1 saddle point representing the ionic configurations.
Following the downhill directions from this index-1 saddle points reveals that it connects the 
covalent local minimum with the lowest-energy ionic local minimum.
Furthermore, the downhill Hessian eigenvector has significant orbital and CI components, which highlights 
the strong coupling between the different degrees of freedom in the vicinity of the avoided crossing.
Both solutions disappear at this point (black dot in Fig.~\ref{fig:LiF_custom}),
and thus there is no quasi-diabatic covalent solution at shorter bond lengths.

In the mathematical framework of catastrophe theory,\cite{ThomBook} this type of coalescence 
can be classified as a fold catastrophe, or a pair annihilation point.
Singularities in this class have previously been identified and characterised for multiple HF solutions,\cite{Fukutome1975}
where they most commonly occur in asymmetric molecules, for example \ce{LiF},\cite{Thom2009,Burton2020} 
\ce{H-Z}\cite{Burton2018} (for a partial nuclear charge $Z$), and ethylene analogues.\cite{Burton2018}
The discontinuous jump in the energy at the pair annihilation point in \ce{LiF} will create issues for 
calculations that attempt to follow the covalent solution across multiple bond lengths, making
these solutions unsuitable for techniques such as \textit{ab initio} molecular dynamics.
Furthermore, since the lowest energy covalent and ionic local minima cross rather than coalesce,
the gradient of the global minimum energy at the crossing point is discontinuous and 
there is an unphysical cusp in the resulting energy surface.

The absence of this pair annihilation point using 6-31G compared to Bauschlicher and Langhoff's basis set
demonstrates how the topology of multiple CASSCF solutions can be affected by the AO basis.
We suspect that these differences arise from the subtle changes in the underlying energy landscape 
that affect the relative stability of different solutions.
However, these results demonstrate the danger of generalising conclusions from one basis set
to another, even for the same molecule.

\section{Concluding Remarks}
\label{sec:conclusion}
State-specific approximations promise to provide a more balanced representation of electronic 
excitations by independently optimising both the ground- and excited-state wave functions.
In this work, we have investigated the energy landscape for excited state-specific stationary points 
in the multi-configurational CASSCF approach.
We have shown how state-specific approximations can accurately describe high-energy and charge transfer
excitations, beyond the reach of state-averaged calculations with small active spaces.
However, the CASSCF energy landscape can have a large number of stationary points, which complicates
the selection and interpretation of physically relevant solutions.

Multiple stationary points in state-specific CASSCF calculations arise through two primary mechanisms.
Firstly, many solutions occur when the active space is too large for the static correlation that 
must be described. 
In this case, the  redundant active orbitals with $n_\text{occ} \approx 0$ can be interchanged with 
virtual orbitals without significantly changing the energy, creating a series of stationary points with 
an increasing number of downhill Hessian eigendirections.
Active orbitals with $n_\text{occ} \approx 2$ can be interchanged with doubly occupied inactive orbitals 
in a similar fashion.
On the other hand, symmetry broken solutions occur when the active space is too small to describe
the static correlation mechanisms, causing the CASSCF wave function to become ``pinned'' 
onto a subset of the  configurations in the exact wave function.
These results demonstrate the importance of finding a ``Goldilocks region'', where the active space
is neither too large or too small, but just right.

Unphysical solutions can have important consequences for the resulting potential energy surfaces.
For example, while choosing the wrong active space only introduces a small energy error when the wave function
is dominated by a single closed-shell configuration,
it can prevent the CASSCF wave function from correctly capturing 
static correlation when the molecular structure changes. 
The active space for stationary points does not change significantly along a reaction coordinate, 
meaning that the incorrect active orbitals remain for all geometries.
For ground-state calculations, one can rely on following downhill directions away from saddle points 
to obtain a more suitable local minimum, hopefully with the best active space.
However, it is hard to predict which Hessian index will give the most physical stationary point for an excited state,
and thus choosing the most accurate excited-state stationary point is challenging 
without prior chemical intuition.
It has long been known that the right choice of active orbitals is key to the success of CASSCF,
but the current results demonstrate the severity of this challenge for state-specific excitations.

In addition, we have investigated the topology of SS-CASSCF\,$(2,2)$ solutions near the singlet-triplet 
conical intersection in \ce{CH2} and the covalent-ionic avoided crossing in \ce{LiF}.
We observe unphysical root flipping where the \ce{CH2} excited state solution is a local minimum near 
the conical intersection, before becoming an index-1 saddle point further along the reaction trajectory.
This phenomenon occurs because the state-specific orbital optimisation artificially stabilises
the local minima, and is still present in the full valence $(6,6)$ active space.
Furthermore, the change in Hessian index is associated with an additional spin-contaminated index-1 saddle point
that connects the singlet and triplet stationary points.
The presence of zero Hessian eigenvalues at these instability thresholds may cause numerical 
issues for second-order optimisation algorithms.
On the other hand, for the \ce{LiF} avoided crossing, we have observed the coalescence of
the local covalent minimum with an index-1 saddle point representing the ionic state, which both
disappear entirely at shorter bond lengths.
While this pairwise coalescence depends on the basis set, it would catastrophically affect
the applicability of SS-CASSCF for generating smooth and continuous potential energy surfaces.

Moving forwards, SS-CASSCF calculations must overcome the troublesome issues of multiple solutions. 
Practical solutions may rely on the identification of suitable initial guesses from more black-box 
techniques, or by focussing on 
optimisation algorithms that target desirable excited-state physical properties 
(e.g.\ dipole moments), such as the generalized variational principles developed by Hanscam and 
Neuscamman.\cite{Hanscam2022}
Alternatively, more bespoke excited-state wave function \textit{ans\"{a}tze}, such as
minimal configuration state functions\cite{Kossoski2022a} 
or excited-state mean-field theory,\cite{Shea2018,Hardikar2020,Shea2020}
may remove unphysical solutions associated with redundant active orbitals and avoid the disappearance
of solutions at pairwise coalescence points.
Surmounting these issues will allow the benefits of state-specific calculations for computing 
excited states, with bespoke orbitals and small active spaces, to be fully realised.

\section*{Supporting Information}
Derivation of the gradient and second-derivatives for the CASSCF energy, 
description of eigenvector-following and Newton--Raphson optimisation algorithms used (PDF).

\section*{Acknowledgements}
H.G.A.B was supported by New College, Oxford, through the Astor Junior Research Fellowship.
The authors thank David Tew for support and computing resources.

\section*{References}
\bibliography{manuscript}

\end{document}